\documentclass[12pt,english,pra,aps,amsmath,amssymb,showpacs,preprint]{revtex4}
\usepackage[T1]{fontenc}
\usepackage[latin9]{inputenc}
\usepackage{amsmath}
\usepackage{amssymb}
\usepackage{esint}

\makeatletter
\@ifundefined{textcolor}{}
{%
 \definecolor{BLACK}{gray}{0}
 \definecolor{WHITE}{gray}{1}
 \definecolor{RED}{rgb}{1,0,0}
 \definecolor{GREEN}{rgb}{0,1,0}
 \definecolor{BLUE}{rgb}{0,0,1}
 \definecolor{CYAN}{cmyk}{1,0,0,0}
 \definecolor{MAGENTA}{cmyk}{0,1,0,0}
 \definecolor{YELLOW}{cmyk}{0,0,1,0}
 }

\usepackage{subfigure}\usepackage{hyperref}\usepackage{float}\hypersetup{
unicode=false,
pdftoolbar=true,
pdfmenubar=true,
pdffitwindow=false,
pdfstartview={FitH},
pdftitle={Mytitle},
pdfauthor={Author},
pdfsubject={Subject},
pdfcreator={Creator},
pdfproducer={Producer},
pdfkeywords={keywords},
pdfnewwindow=true,
colorlinks=true,
linkcolor=red,
citecolor=blue,
filecolor=magenta,
urlcolor=cyan
}

\makeatother

\usepackage{babel}

\makeatother

\usepackage{babel}

\makeatother

\usepackage{babel}

\makeatother

\usepackage{babel}

\makeatother

\usepackage{babel}

\makeatother

\usepackage{babel}

\makeatother

\usepackage{babel}

\makeatother

\usepackage{babel}

\makeatother

\usepackage{babel}

\begin{document}

\title{A new quasi-exactly solvable problem and its connection with an anharmonic
oscillator}

\author{Da-Bao Yang}

\author{Fu-Lin Zhang}

\author{Jing-Ling Chen}

\email{chenjl@nankai.edu.cn}

\affiliation{Chern Institute of Mathematics, Nankai University, Tianjin 300071,
China}

\date{\today}
\begin{abstract}
The two-dimensional hydrogen with a linear potential in a magnetic
field is solved by two different methods. Furthermore the connection
between the model and an anharmonic oscillator is investigated by
methods of KS transformation. 
\end{abstract}
\maketitle
\setlength{\baselineskip}{20pt}

\section{Introduction}

The discovery of quasi-exactly solvable quantal problems was a remarkable
contribution in last century\cite{shifman1989new}. This kind of problem
could be solved by lie algebraic methods, which was pioneered by Turbiner\cite{Turbiner1988quasi}.
The method was also be applied to an anharmonic oscillators by Chen
et. al.\cite{chen2001solving}. Moreover a generalization of the basic
lie algebraic construction was considered by Shifman\cite{shifman1989expanding}.
You can also confer to a reviewed paper by Shifman\cite{shifman1989new}.
However, the quasi-exactly solvable problem could also be calculated
by an analytical way\cite{Taut1995Two}.

This paper is concentrated on a model that is a two-dimensional hydrogen
with a linear potential in a homogeneous magnetic field. This model
is very meaningful among the quasi-exactly problems. First, its potential
is novel and complex, moreover when the coefficient of the linear
potential becomes zero, its all result could go back to the simpler
circumstance which have been investigated by Taut\cite{Taut1995Two}.
Second, the sl(2) algebra structure hidden in the Hamiltonian is not
apparent, however it will be revealed in this paper. Third, this model
is also solved by an analytical method whose result coincide with
the one by the sl(2) method. Last but not least, a new sextic oscillator
with a centrifugal potential is obtained from this model. Furthermore,
the connection between them is constructed.

This paper is organised as follows: in section two, the separation
of variables in the corresponding schr$\ddot{o}$dinger equation is
displayed, so we get the radial part equation; in section three, we
solve it by the lie-algebraic methods, i.e., sl(2); in section four,
it is solved by an analytic way, therefore, the two results coincide
with each other; in section five we build its connection with the
sextic oscillator with a centrifugal barrier; in section six, a conclusion
is drawed.

\section{The separation of variables}

\label{sec:system}

Now, we consider a two-dimensional hydrogen with a linear potential
in a homogeneous magnetic field, where the magnetic field $\vec{B}$
is perpendicular to the plan in which the electron is located. The
schr$\ddot{o}$dinger equation reads

\[
[\frac{1}{2}(\vec{p}+\frac{1}{c}\vec{\mathcal{A}})^{2}-\frac{Z}{r}+kr]\psi=E\psi~~,\]
 where $c$ is the velocity of light and the vector potential in the
symmetric gauge is $\vec{\mathcal{A}}=\frac{1}{2}\vec{B}\times\vec{r}.$
Under some operation, the above equation can be transformed to be

\begin{equation}
\frac{1}{2}[\vec{p}~^{2}\psi+\frac{1}{c}(\vec{p}\cdot\vec{\mathcal{A}})\psi+\frac{2}{c}\vec{\mathcal{A}}\cdot\vec{p}\psi+\frac{1}{c^{2}}\vec{\mathcal{A}}^{2}\psi]-\frac{Z}{r}\psi+kr\psi=E\psi~~,\label{eq:BeforeTransformedSch}\end{equation}
 In polar coordinates, letting $r$ and $\theta$ represent the radius
and azimuthal angle respectively, we can get\[
\begin{cases}
\vec{p}=-i(\vec{e_{r}}\frac{\partial}{\partial r}+\vec{e_{\theta}}\frac{1}{r}\frac{\partial}{\partial\theta})\\
\vec{\mathcal{A}}=\frac{1}{2}\vec{B}\times\vec{r}=\frac{1}{2}Br\vec{e_{\theta}}\\
\vec{p}\cdot\vec{\mathcal{A}}=0\\
\vec{\mathcal{A}}\cdot\vec{p}=-i\frac{1}{2}B\frac{\partial}{\partial\theta}\\
\vec{\mathcal{A}}^{2}=\frac{1}{4}B^{2}r^{2}\end{cases}.\]
 Substitution the above equations into the Eq.\eqref{eq:BeforeTransformedSch},
we get

\begin{equation}
\frac{1}{2}[-\nabla^{2}\psi-i2\omega_{L}\frac{\partial}{\partial\theta}\psi+\omega_{L}^{2}r^{2}\psi]-\frac{Z}{r}\psi+kr\psi=E\psi,\label{eq:TransformedSch}\end{equation}
 where $\nabla^{2}=\frac{1}{r}\frac{\partial}{\partial r}r\frac{\partial}{\partial r}+\frac{1}{r^{2}}\frac{\partial}{\partial\theta^{2}}$
and $\omega_{L}=\frac{B}{2c}$. Having represented the equation in
a concrete coordinate, we will separate the variables in the next
paragraph.

Substituting \begin{equation}
\psi(r,\theta)=\frac{1}{\sqrt{2\pi}}e^{im\theta}R(r)\label{eq:TotalWaveFunction}\end{equation}
 into the above equation, where $m=0,\pm1,\pm2,...$, we get the radial
part equation, which is \begin{equation}
(-\frac{1}{2}\frac{d^{2}}{dr^{2}}-\frac{1}{2}\frac{1}{r}\frac{d}{dr}+\frac{\omega_{L}^{2}r^{2}}{2}+kr+\omega_{L}m-\frac{Z}{r}+\frac{m^{2}}{2r^{2}})R=ER.\label{eq:radialsch}\end{equation}
 The above differential equation will be solved by the method of sl(2)
algebra in the next secion.

\section{Solving the radial equation with the sl(2) algebra}

\label{sec:sl(2)}

First of all, let us recall the essential idea that the schr$\ddot{o}$dinger
equation is solved by sl(2). The schr$\ddot{o}$dinger equation determining
the stationary states of the system is

\[
H\psi(r)=E\psi(r),\]

\[
H=-\frac{1}{2}\frac{d^{2}}{dr^{2}}+V(r).\]
 According to \cite{shifman1989new}, we can perform a quasi-gauge
transformation

\begin{equation}
\psi(r)=\tilde{\psi}(r)exp(-\int{A(r)dr}).\label{eq:wavefunction}\end{equation}
 Then the stationary schr$\ddot{o}$dinger equation can be transformed
to\[
H_{"G"}\tilde{\psi}=E\tilde{\psi},\]
 \[
H_{"G"}=-\frac{1}{2}(\frac{d}{dx}-A(x))^{2}+V(x).\]
 Furthermore $H_{"G"}$ can be represented by the combination of the
partial algebraization of the generators of sl(2), which is

\begin{equation}
H_{"G"}=\sum_{a,b=\pm,0}C_{ab}T^{a}T^{b}+\sum_{a=\pm,0}C_{a}T^{a}+const,\label{eq:combination}\end{equation}
 where

\[
T^{+}=2jr-r^{2}\frac{d}{dr},\]

\begin{equation}
T^{0}=-j+r\frac{d}{dr},\label{eq:generators}\end{equation}

\[
T^{-}=\frac{d}{dr}.\]

It is very easy to check that the commutation relations for $T^{\pm}$
and $T^{0}$ are indeed those inherent to the sl(2) algebra:

\[
[T^{+},T^{-}]=2T^{0},~~~~~[T^{0},T^{+}]=+T^{+},~~~~~[T^{0},T^{-}]=-T^{-}.\]
 Finally, with the aid of the above commutation relations, the transformed
Hamiltonian $H_{"G"}$ can be represented by a $(2j+1)\times(2j+1)$
matrix, whose basis are\[
\{\tilde{\psi}\}=\{1,r,r^{2},\cdots,r^{2j-1},r^{2j}\}.\]
 Then the standard algorithm of diagonalization of the finite matrix
gives us $(2j+1)$ eigenvalues and the corresponding eigenfunctions
of the transformed Hamiltonian.

Blindly following the program outlined above, we do not find the above
model that we pay attention to in the section 2 can be solved by the
methods of sl(2), in another word, the Hamiltonian $H$ isn't reducible
to the quadratic combination of the sl(2) generators. However, we
can consider a corresponding eigenvalue problem, $f(r)H\psi=E\psi$,
where $f$ is an appropriately chosen function of the variable $r$,
so that $fH$ can be reducible to the standard form \eqref{eq:combination}.
The above explanation maybe a litter obscure, moreover this method
will be showed in more transparent terms in the following paragraphs.

Let us introduce an auxiliary Hamiltonian\begin{equation}
\tilde{H}=\gamma(-\frac{1}{2}\frac{d^{2}}{dr^{2}}-\frac{1}{2r}\frac{d}{dr}+\frac{\omega_{L}^{2}r^{2}}{2}+\omega_{L}m+\frac{m^{2}}{2r^{2}})-\frac{Z}{r}+\gamma kr-\gamma X,\label{eq:auxiliaryH-1}\end{equation}
 where $\gamma$ and $X$ are constant coefficients.Next, we calculate
the {}``gauge''-transformed Hamiltonian $\tilde{H}_{"G"}$, simply
stemming form $\tilde{H}$ after the substitution\begin{equation}
\frac{d}{dr}\rightarrow\frac{d}{dr}-A.\label{eq:transformation}\end{equation}
 Then\begin{equation}
\tilde{H}_{"G"}=\gamma(-\frac{1}{2}\frac{d^{2}}{dr^{2}}-\frac{1}{2r}\frac{d}{dr}+A\frac{d}{dr})-\frac{1}{2}\gamma A^{2}+\gamma\frac{\omega_{L}^{2}r^{2}}{2}+\gamma kr+\frac{1}{2}\gamma A'+\gamma\frac{A}{2r}+\gamma\frac{m^{2}}{2r^{2}}-\frac{Z}{r}+\gamma(m\omega_{L}-X).\label{eq:auxiliaryH}\end{equation}
 After a short analysis, we can get a conclusion that: when $r\rightarrow\infty$,
the term $\gamma\frac{\omega_{L}^{2}r^{2}}{2}+\gamma kr$ will become
the dominant one; when $r\rightarrow0$, the term $\gamma\frac{m^{2}}{2r^{2}}$
will become the dominant counterpart. In order to guarantee the normalizability
of the wave function, we must choose a proper form of $A$ to cancel
out the above terms. A short reflection shows that\begin{equation}
A=\mu r+\delta+\frac{\nu}{r}.\label{eq:A}\end{equation}
 Substituting the above Eq.\eqref{eq:A} into Eq.\eqref{eq:auxiliaryH},
one gets

\begin{equation}
\begin{array}{ccl}
\tilde{H}_{"G"} & = & \gamma(-\frac{1}{2}\frac{d^{2}}{dr^{2}}-\frac{1}{2r}\frac{d}{dr}+A\frac{d}{dr})+\frac{1}{2}\gamma(\omega_{L}^{2}-\mu^{2})r^{2}+\gamma(k-\mu\delta)r+(\frac{1}{2}\gamma\delta-\gamma\delta\nu-Z)\frac{1}{r}\\
 &  & +\frac{1}{2}\gamma(m^{2}-\nu^{2})\frac{1}{r^{2}}+\gamma(+\omega_{L}m+\mu-\mu\nu-\frac{1}{2}\delta^{2}-X)\end{array}.\label{eq:auxiliaryH1}\end{equation}
 Moreover, from above analysis, one knows the coefficients of $r^{2}$,
$r$ and $\frac{1}{r^{2}}$ are equal to $0$ respectively, which
are \[
\begin{cases}
\omega_{L}^{2}-\mu^{2}=0\\
k-\mu\delta=0\\
m^{2}-\nu^{2}=0\end{cases}\]
 The above equations can be solved out:\begin{equation}
\begin{cases}
\mu=\pm\omega_{L}\\
\delta=\frac{k}{\mu}\\
\nu=\pm|m|\end{cases}.\label{eq:coefficients}\end{equation}
 Substituting Eq.\eqref{eq:A} and Eq.\eqref{eq:coefficients} into
Eq.\eqref{eq:wavefunctionFinal}, one finds that only\begin{equation}
R(r)=\tilde{R(r)}\exp(-\frac{1}{2}\omega_{L}r^{2}-\frac{k}{\omega_{L}}r+|m|\ln r)\label{eq:wavefunctionFinal}\end{equation}
 can guarantee that the wave function converge at both sides of $r\rightarrow0$
and $r\rightarrow\infty$(Note that we replace $\psi(r)$ and $\tilde{\psi}(r)$
by $R(r)$ and $\tilde{R}(r)$ respectively.), that is to say\begin{equation}
\begin{cases}
\mu=+\omega_{L}\\
\delta=\frac{k}{\omega_{L}}\\
\nu=-|m|\end{cases}.\label{eq:coefficients1}\end{equation}
 Substituting Eq.\eqref{eq:coefficients1} into Eq.\eqref{eq:auxiliaryH1},
one can gets\begin{equation}
\frac{r}{\gamma}\tilde{H}_{"G"}\tilde{R}=\frac{E(\gamma)}{\gamma}\tilde{R},\label{eq:transformedsch}\end{equation}
 where\[
\begin{array}{c}
\tilde{H}{}_{"G"}=\gamma[-\frac{1}{2}\frac{d^{2}}{dr^{2}}-\frac{1}{2r}\frac{d}{dr}+(\omega_{L}r+\frac{k}{\omega_{L}}-\frac{|m|}{r})\frac{d}{dr}]\\
+[\gamma\frac{k}{\omega_{L}}(|m|+\frac{1}{2})-Z]+\gamma[\omega_{L}(1+m+|m|)-\frac{1}{2}(\frac{k}{\omega_{L}})^{2}-X].\end{array}\]
 The reason that I divided both sides of Eq.\eqref{eq:transformedsch}
by $\gamma$ is for a convenient calculation.The reader maybe puzzled
by the reason why we use $r$ to multiple $\tilde{H}_{"G"}$, because
$r\tilde{H}_{"G"}$ can reducible to a quadratic combination of the
generator of sl(2).Under the inverse transformation Eq.\eqref{eq:transformation},
the Eq.\eqref{eq:transformedsch} can be transformed to be

\begin{equation}
\frac{1}{\gamma}\tilde{H}R(r)=\frac{E(\gamma)}{\gamma r}R(r),\label{eq:transformedsch1}\end{equation}
 where \[
R=\tilde{R}\exp(-\frac{1}{2}\omega_{L}r^{2}-\frac{k}{\omega_{L}}r+|m|\ln r).\]
 Substituting Eq.\eqref{eq:auxiliaryH-1} into Eq.\eqref{eq:transformedsch1},
we can get\begin{equation}
(-\frac{1}{2}\frac{d^{2}}{dr^{2}}-\frac{1}{2r}\frac{d}{dr}+\frac{\omega_{L}^{2}r^{2}}{2}+kr+\omega_{L}m-\frac{1}{\gamma}\frac{Z+E}{r}+\frac{m^{2}}{2r^{2}})R=XR.\label{eq:radicalsch1}\end{equation}
 By comparison with the Eq.\eqref{eq:radialsch}, we observe that
we have actually solved the original problem provided that the parameter
$\gamma$ is chosen as follows\begin{equation}
\frac{Z+E}{\gamma}=Z.\label{eq:condition}\end{equation}
 And we represented the eigenvalue by another symbol $X$ instead.

In order to solve Eq.\eqref{eq:radicalsch1} by sl(2), let's return
to Eq.\eqref{eq:transformedsch}. Following what I said before, the
$\frac{r}{\gamma}\tilde{H}_{"G"}$ can be reducible to the combination
of generators of sl(2), which is\begin{equation}
\frac{r}{\gamma}\tilde{H}_{"G"}=C_{1}T^{0}T^{-}+C_{2}T^{+}+C_{3}T^{0}+C_{4}T^{-}+C_{0}.\label{eq:combination1}\end{equation}
 Substituting Eq.\eqref{eq:generators} into the above Eq.\eqref{eq:combination1},
we can get\[
\frac{r}{\gamma}\tilde{H}_{"G"}=C_{1}r\frac{d^{2}}{dr^{2}}-C_{2}r^{2}\frac{d}{dr}+C_{3}r\frac{d}{dr}+(C_{4}-jC_{1})\frac{d}{dr}+2C_{2}jr-jC_{3+}C_{0}.\]
 By comparison with Eq.\eqref{eq:transformedsch}, the coefficients
can be solved from equations\[
\begin{cases}
C_{1}=-\frac{1}{2}\\
-C_{2}=\omega_{L}\\
C_{3}=\frac{k}{\omega_{L}}\\
C_{4}-jC_{1}=-|m|-\frac{1}{2}\\
2C_{2}j=\omega_{L}(1+m+|m|)-X\\
-jC_{2}+C_{0}=\frac{k}{\omega_{L}}(\frac{1}{2}+|m|)-\frac{Z}{r}\end{cases}.\]
 A further simplified calculation show that\begin{equation}
\begin{cases}
C_{1}=-\frac{1}{2}\\
C_{2}=-\omega_{L}\\
C_{3}=\frac{k}{\omega_{L}}\\
C_{4}=-\frac{1}{2}(1+j+2|m|)\\
X=\omega_{L}(2j+1+m+|m|)-\frac{1}{2}(\frac{k}{\omega_{L}})^{2}\\
C_{0}=\frac{k}{\omega_{L}}(|m|+\frac{1}{2}+j)-\frac{Z}{\gamma}\end{cases}.\label{eq:coefficientsC}\end{equation}
 Substituting the above Eq.\eqref{eq:coefficientsC} and Eq.\eqref{eq:combination1}
into Eq.\eqref{eq:transformedsch}, one gets\begin{equation}
[-\frac{1}{2}T^{0}T^{-}-\omega_{L}T^{+}+\frac{k}{\omega_{L}}T^{0}-\frac{1}{2}(1+j+|m|)T^{-}+\frac{k}{\omega_{L}}(|m|+\frac{1}{2}+j)]\tilde{R}=\frac{Z+E}{\gamma}\tilde{R}\label{eq:transformedsch2}\end{equation}
 By use of the constraint which is Eq.\eqref{eq:condition}, the above
Eq.\eqref{eq:transformedsch2} can be transformed to\begin{equation}
[-\frac{1}{2}T^{0}T^{-}-\omega_{L}T^{+}+\frac{k}{\omega_{L}}T^{0}-\frac{1}{2}(1+j+|m|)T^{-}+\frac{k}{\omega_{L}}(|m|+\frac{1}{2}+j)]\tilde{R}=Z\tilde{R}\label{eq:transformedschFinal}\end{equation}

1)j=0

Hence, we know \begin{equation}
\tilde{R}=a_{0}.\label{eq:r0}\end{equation}
 Substituting Eq.\eqref{eq:r0} into Eq.\eqref{eq:wavefunctionFinal},
one get \begin{equation}
R=a_{0}\exp(-\frac{1}{2}\omega_{L}r^{2}-\frac{k}{\omega_{L}}r+|m|\ln r).\label{eq:RadialWaveFunctionJZero}\end{equation}
 Substituting Eq.\eqref{eq:generators} and Eq.\eqref{eq:r0} into
Eq.\eqref{eq:transformedschFinal}, we can get\begin{equation}
\frac{k}{\omega_{L}}(|m|+\frac{1}{2})=Z,\label{eq:eigenvalue0}\end{equation}
 which is precisely the constraint of eigenvalues. Substituting the
above Eq.\eqref{eq:eigenvalue0} into the fifth Equation of Eq.\eqref{eq:coefficientsC},
we can get the eigenvalue\[
X_{0}=\omega_{L}(1+m+|m|)-\frac{1}{2}(\frac{Z}{|m|+\frac{1}{2}})^{2}\]

2)$j=\frac{1}{2}$

It is very easy to calculate that\begin{equation}
T^{+}=\left(\begin{array}{cc}
0 & 1\\
0 & 0\end{array}\right)~~~~~T^{0}=\frac{1}{2}\left(\begin{array}{cc}
1 & 0\\
0 & -1\end{array}\right)~~~~~T^{-}=\left(\begin{array}{cc}
0 & 0\\
1 & 0\end{array}\right).\label{eq:GeneratorsOneSecond}\end{equation}
 Substitution the above Eq.\eqref{eq:GeneratorsOneSecond} into Eq.\eqref{eq:transformedschFinal},
we can get\begin{equation}
\left(\begin{array}{cc}
\frac{k}{\omega_{L}}(|m|+\frac{3}{2}) & -\omega_{L}\\
-(|m|+\frac{1}{2}) & \frac{k}{\omega_{L}}(|m|+\frac{1}{2})\end{array}\right)\left(\begin{array}{c}
a_{1}\\
a_{0}\end{array}\right)=Z\left(\begin{array}{c}
a_{1}\\
a_{0}\end{array}\right).\label{eq:EigenEqationOneSecond}\end{equation}
 Such a equation has a nontrivial solution if the determinant of the
corresponding matrix vanishes, i.e. \[
\left|\begin{array}{cc}
\frac{k}{\omega_{L}}(|m|+\frac{3}{2})-Z & -\omega_{L}\\
-(|m|+\frac{1}{2}) & \frac{k}{\omega_{L}}(|m|+\frac{1}{2})-Z\end{array}\right|=0.\]
 The equation can be further simplified to be\begin{equation}
\omega_{L}^{2}Z^{2}-2k\omega_{L}(|m|+1)Z+k^{2}(|m|+\frac{3}{2})(|m|+\frac{1}{2})-\omega_{L}^{3}(|m|+\frac{1}{2})]=0.\label{eq:ConstraintOneSecond}\end{equation}
 Form it, we can draw a conclusion that the parameters $\omega_{L}$
and $k$ can't be chosen arbitrarily, they must satisfy the above
Eq.\eqref{eq:ConstraintOneSecond}. The above Eq.\eqref{eq:ConstraintOneSecond}
can be regarded as a quadratic Equation about $Z$. As a result, the
solution of it can be expressed as\begin{equation}
Z=\frac{k(|m|+1)\pm\sqrt{M}}{\omega_{L}},\label{eq:RootZ}\end{equation}
 where $M\equiv\frac{k^{2}}{4}+\omega_{L}^{3}(\frac{1}{2}+|m|)$.
In the following paragraph, we will calculate the different eigenfunctions
corresponding to different eigenvalues.

a)$Z=\frac{k(|m|+1)+\sqrt{M}}{\omega_{L}},$

Substituting the above equation into Eq.\eqref{eq:EigenEqationOneSecond},
we get\begin{equation}
a_{0}\left(\begin{array}{c}
-\frac{\frac{1}{2}\frac{k}{\omega_{L}}+\frac{1}{\omega_{L}}\sqrt{M}}{\frac{1}{2}+|m|}\\
1\end{array}\right)\equiv a_{0}(1-\frac{\frac{1}{2}\frac{k}{\omega_{L}}+\frac{1}{\omega_{L}}\sqrt{M}}{\frac{1}{2}+|m|}r).\label{eq:PolynomialsOneSecond}\end{equation}
 Substituting the above Eq.\eqref{eq:PolynomialsOneSecond} into Eq.\eqref{eq:wavefunctionFinal}\begin{equation}
R(r)_{j=\frac{1}{2}}^{m_{j}=\frac{1}{2}}=a_{0}(1-\frac{\frac{1}{2}\frac{k}{\omega_{L}}+\frac{1}{\omega_{L}}\sqrt{M}}{\frac{1}{2}+|m|}r)\exp(-\frac{1}{2}\omega_{L}r^{2}-\frac{k}{\omega_{L}}r+|m|\ln r).\label{eq:RadialPartWaveFunctionJOneSecondPositive}\end{equation}

b)$Z=\frac{k(|m|+1)-\sqrt{M}}{\omega_{L}},$

Following the above procedure, we get the radial part of the wave
function, i.e.,\begin{equation}
R(r)_{j=\frac{1}{2}}^{m_{j}=-\frac{1}{2}}=a_{0}(1-\frac{\frac{1}{2}\frac{k}{\omega_{L}}-\frac{1}{\omega_{L}}\sqrt{M}}{\frac{1}{2}+|m|}r)\exp(-\frac{1}{2}\omega_{L}r^{2}-\frac{k}{\omega_{L}}r+|m|\ln r).\label{eq:RadialPartWaveFunctionJOneSecondNegative}\end{equation}
 With the constraint of Eq.\eqref{eq:ConstraintOneSecond}, the eigenvalue
of energy becomes\[
X_{\frac{1}{2}}=\omega_{L}(2+m+|m|)-\frac{1}{2}(\frac{k}{\omega_{L}})^{2}\]

3)Arbitrary j

According to the choice of the phase, we know the following formulation,
i.e.,

\[
T^{\pm}|jm\rangle=\sqrt{(j\mp m)(j\pm m+1)}|j,m\pm1\rangle~~~~~T^{0}|jm\rangle=m|jm\rangle.\]
 The $j_{\pm}$ and $j_{0}$ can be represented in the following $(2j+1)\times(2j+1)$
matrix,

\begin{equation}
T^{+}=\left(\begin{array}{cccccc}
0 & \sqrt{1\cdot2j} & 0 & ~ & 0 & 0\\
0 & 0 & \sqrt{2\cdot(2j-1)} & ~ & 0 & 0\\
\vdots & \vdots & \vdots & \ddots & \vdots & \vdots\\
0 & 0 & 0 & \cdots & \sqrt{(2j-1)\cdot2} & 0\\
0 & 0 & 0 & ~ & 0 & \sqrt{2j\cdot1}\\
0 & 0 & 0 & ~ & 0 & 0\end{array}\right)\label{eq:creation}\end{equation}

\begin{equation}
T^{-}=\left(\begin{array}{cccccc}
0 & 0 & ~ & 0 & 0 & 0\\
\sqrt{1\cdot2j} & 0 & ~ & 0 & 0 & 0\\
0 & \sqrt{2\cdot(2j-1)} & ~ & 0 & 0 & 0\\
\vdots & \vdots & \ddots & \vdots & \vdots & \vdots\\
0 & 0 & ~ & \sqrt{(2j-1)\cdot2} & 0 & 0\\
0 & 0 & ~ & 0 & \sqrt{2j\cdot1} & 0\end{array}\right)\label{eq:annihilation}\end{equation}

\begin{equation}
T^{0}=\left(\begin{array}{cccccc}
j & ~ & ~ & ~ & ~ & ~\\
~ & j-1 & ~ & ~ & ~ & ~\\
~ & ~ & j-2 & ~ & ~ & ~\\
~ & ~ & ~ & \ddots & ~ & ~\\
~ & ~ & ~ & ~ & -j+1 & ~\\
~ & ~ & ~ & ~ & ~ & -j\end{array}\right)\label{eq:weight}\end{equation}
 Substitution the above Eq.\eqref{eq:creation}, Eq.\eqref{eq:annihilation}
and Eq.\eqref{eq:weight} into Eq.\eqref{eq:transformedschFinal},
so the problem is deduced to an eigenvalue and eigenfunction problem.
In order to get the nontrivial eigenfunction, the determinant of the
corresponding matrix must vanish, which will give a onstraint about
$\omega_{L}$ and $k$.Similarly, we will get the radial part of the
wave function. Furthermore, the fifth equation of Eq.\eqref{eq:coefficientsC}
tell us the eigenvalue of energy,\[
X_{j}=\omega_{L}(2j+1+m+|m|)-\frac{1}{2}(\frac{k}{\omega_{L}})^{2}.\]
 To sum up, in this section, we have solved Eq.\eqref{eq:radialsch}
by method of sl(2). Moreover, we will solved it again by another method,
namely, an analytical methods.

\section{Solving the same equation with analytical method}

In order to solve Eq.\eqref{eq:radialsch}, using the following transformation,
i.e.\begin{equation}
R=\frac{u}{\sqrt{r}},\label{eq:TransformationRU}\end{equation}
 we can get\begin{equation}
-\frac{1}{2}\frac{d^{2}u}{dr^{2}}+\frac{1}{2}(m^{2}-\frac{1}{4})\frac{1}{r^{2}}u-\frac{Z}{r}u+\omega_{L}mu+kru+\frac{1}{2}\omega_{L}^{2}r^{2}u=Eu.\label{eq:TransformedRadialSch}\end{equation}
 Due to the condition of square-integrability of the wave function,
it is useful to consider of the limits $r\rightarrow0$ and $r\rightarrow\infty$
of the above Eq.\eqref{eq:TransformedRadialSch}. For $r\rightarrow0$,
the term containing $\frac{1}{r^{2}}$ is dominant and we get the
equation\begin{equation}
-\frac{1}{2}\frac{d^{2}u}{dr^{2}}+\frac{1}{2}(m^{2}-\frac{1}{4})\frac{1}{r^{2}}u=0.\label{eq:LimitZero}\end{equation}
 Because the above Eq.\eqref{eq:LimitZero} contains a normal singularity,
so we substitute a power series $u=r^{s}\sum_{i=0}^{\infty}b_{i}r^{i}$
into the equation and one gets\[
-\sum_{i=0}^{\infty}(s+i)(s+i-1)b_{i}r^{s+i-2}+(m^{2}-\frac{1}{4})\sum_{i=0}^{\infty}b_{i}r^{s+i-2}=0.\]
 Due to the uniqueness of the series, the coefficient of it is equal
to zero, we can get\[
-(s+i)(s+i-1)b_{i}+(m^{2}-\frac{1}{4})b_{i}=0.\]
 By use of the limits $r\rightarrow0$, only the lowest-order term
become dominant, so we omits the higher order terms and get\[
-s(s-1)b_{0}+(m^{2}-\frac{1}{4})b_{0}=0.\]
 In order to get the nontrivial solution, we must set $b_{0}\neq0$,
so one gets $s=m+\frac{1}{2}$ or $s=-m+\frac{1}{2}$. However, if
m become negative, the solution diverges when $r\rightarrow0$. Hence,
we get\[
s=|m|+\frac{1}{2}.\]
 So the solution of Eq.\eqref{eq:LimitZero} is\begin{equation}
u=r^{|m|+\frac{1}{2}}.\label{eq:SolutionZero}\end{equation}

Having solved the solution $r\rightarrow0$, we consider about another
situation $r\rightarrow\infty$. Under this circumstance, the terms
containing $r^{2}$ and $r$ in Eq.\eqref{eq:TransformedRadialSch}
become dominant. So the equation becomes\[
-\frac{1}{2}\frac{d^{2}u}{dr^{2}}+\frac{1}{2}\omega_{L}^{2}r^{2}u+kru=0.\]
 We can verify the solution of the above equation is\begin{equation}
u=\exp(-\frac{1}{2}\omega_{L}r^{2}-\frac{k}{\omega_{L}}r).\label{eq:SolutionInfinity}\end{equation}
 Considering both Eq.\eqref{eq:SolutionZero} and Eq.\eqref{eq:SolutionInfinity},
we can write the solution of Eq.\eqref{eq:TransformedRadialSch}as\begin{equation}
u=r^{|m|+\frac{1}{2}}e^{-\frac{1}{2}\omega_{L}r^{2}-\frac{k}{\omega_{L}}r}\varphi.\label{eq:SolutionOfRadialSch}\end{equation}
 Substituting the above Eq.\eqref{eq:SolutionOfRadialSch}into \eqref{eq:TransformedRadialSch},
we get\begin{equation}
\begin{array}{c}
-\frac{1}{2}r\frac{d^{2}\varphi}{dr^{2}}+\omega_{L}r^{2}\frac{d\varphi}{dr}+\frac{k}{\omega_{L}}r\frac{d\varphi}{dr}-(|m|+\frac{1}{2})\frac{d\varphi}{dr}+(|m|\omega_{L}+\omega_{L}+m\omega_{L}-\frac{1}{2}\frac{k^{2}}{\omega_{L}^{2}}-E)r\varphi\\
+[(|m|+\frac{1}{2})\frac{k}{\omega_{L}}-Z]\varphi=0.\end{array}\label{eq:FinalTransformedRadialSch}\end{equation}
 The above differential equation has two nonessential singularities
at $r=0,\infty$. It's index equation takes this form\[
s(s+2|m|)=0.\]
 It's solutions are $0$ and $-2|m|$ respectively. However, if $s=-2|m|$,
the solution diverges, in another words, it violate the demand of
square-integrability of the wave function. Hence, $s$ must be $0$.
Therefore, we can substitute the power series\begin{equation}
\varphi=\sum_{n=0}^{\infty}a_{n}r^{n}\label{eq:SeriesN}\end{equation}
 into Eq.\eqref{eq:FinalTransformedRadialSch}and find the recurrence
relation\[
\begin{array}{c}
-\frac{1}{2}\sum_{n=1}^{\infty}a_{n+1}(n+1)nr^{n}+\omega_{L}\sum_{n=2}^{\infty}a_{n-1}(n-1)r^{n}+\frac{k}{\omega_{L}}\sum_{n=1}^{\infty}a_{n}nr^{n}\\
-(|m|+\frac{1}{2})\sum_{n=0}^{\infty}a_{n+1}(n+1)r^{n}+(|m|\omega_{L}+\omega_{L}+m\omega_{L}-\frac{1}{2}\frac{k^{2}}{\omega_{L}^{2}}-E)\sum_{n=1}^{\infty}a_{n-1}r^{n}\\
+[(|m|+\frac{1}{2})\frac{k}{\omega_{L}}-Z]\sum_{n=0}^{\infty}a_{n}r^{n}=0\end{array}\]
 Let us discuss in detail in the following paragraphs.

1)$n=0$

\begin{equation}
a_{1}=\frac{(|m|+\frac{1}{2})\frac{k}{\omega_{L}}-Z}{|m|+\frac{1}{2}}a_{0}.\label{eq:nZero}\end{equation}
 We put forward a hypothesis that the ground state exists, so we get
$a_{0}\neq0$, $a_{1}=0$, $a_{2}=0$, $\cdots$. However, conferring
to Eq.\eqref{eq:nOne}, if $a_{0}\neq0$, $a_{2}$ may be not equal
to $0$, which is contradict the hypothesis. So we can draw a conclusion
that $a_{0}=0$, that is to say, there isn't a ground state.

2)$n=1$\begin{equation}
a_{2}=\frac{|m|\omega_{L}+\omega_{L}+m\omega_{L}-\frac{1}{2}\frac{k^{2}}{\omega_{L}^{2}}-E}{2(|m|+1)}a_{0}+\frac{(|m|+\frac{3}{2})\frac{k}{\omega_{L}}-Z}{2(|m|+1)}a_{1}.\label{eq:nOne}\end{equation}
 If the first excited state exists, $a_{0}$ and $a_{1}$ can't be
equal to $0$ simultaneously, $a_{2}=0$, $a_{3}=0$, $\cdots$. Conferring
to Eq.\eqref{eq:nSecond}, $a_{1}$ will have a effect on $a_{3}$,
so $a_{1}=0$, that is to say, conferring to Eq.\eqref{eq:nZero},
\[
Z=(|m|+\frac{1}{2})\frac{k}{\omega_{L}}.\]
 So the radial part of wave function is\begin{equation}
R_{n=1}=r^{|m|}e^{-\frac{1}{2}\omega_{L}r^{2}-\frac{k}{\omega_{L}}r}a_{0},\label{eq:RadialWaveFunctionNOne}\end{equation}
 which is gained by use of Eq.\eqref{eq:TransformationRU} and Eq.\eqref{eq:SolutionOfRadialSch}.By
comparison with Eq.\eqref{eq:RadialWaveFunctionJZero}, it is found
easily that the two results coincide with each other. From above analysis,
we can draw a conclusion that $a_{0}\neq0$, $a_{1}=0$, $a_{2}=0$,
$\cdots$, conferring to Eq.\eqref{eq:nOne}, so we get\begin{equation}
E_{1}=\omega_{L}(1+|m|+m)-\frac{1}{2}\frac{k^{2}}{\omega_{L}^{2}}.\label{eq:EnergyOne}\end{equation}

3)$n=2$\begin{equation}
a_{3}=\frac{|m|\omega_{L}+2\omega_{L}+m\omega_{L}-\frac{1}{2}\frac{k^{2}}{\omega_{L}^{2}}-E}{3(|m|+\frac{3}{2})}a_{1}+\frac{(|m|+\frac{5}{2})\frac{k}{\omega_{L}}-Z}{3(|m|+\frac{3}{2})}a_{2}.\label{eq:nSecond}\end{equation}
 Following the above analysis, if the second excited state exists,
we can get $a_{3}=0$, $a_{4}=0$ and so on. It is necessary to demand
the first term and the second term are equal to $0$ respectively,
conferring to Eq.\eqref{eq:nSecond}. Moreover if $a_{2}\neq0$, it
may render $a_{4}\neq0$, which is contradict our hypothesis. So we
must let $a_{2}=0$, in another word, the second term become zero
already. Now let us concentrate on the first term. If $a_{1}=0$,
this case will be the same with the first excited state. So we can
get\begin{equation}
E_{2}=(2+|m|+m)\omega_{L}-\frac{1}{2}\frac{k^{2}}{\omega_{L}^{2}}.\label{eq:EnergyTwo}\end{equation}
 Substituting Eq.\eqref{eq:nZero} and Eq.\eqref{eq:EnergyTwo} into
Eq.\eqref{eq:nOne}, we can get\[
a_{2}=\frac{[(|m|+\frac{3}{2})\frac{k}{\omega_{L}}-Z][(|m|+\frac{1}{2})\frac{k}{\omega_{L}}-Z]-(|m|+\frac{1}{2})\omega_{L}}{2(|m|+1)(|m|+\frac{1}{2})}a_{0}.\]
 From above analysis, we know $a_{2}=0$, that is to say,\[
[(|m|+\frac{3}{2})\frac{k}{\omega_{L}}-Z][(|m|+\frac{1}{2})\frac{k}{\omega_{L}}-Z]-(|m|+\frac{1}{2})\omega_{L}=0,\]
 which coincide with Eq.\eqref{eq:ConstraintOneSecond}. Therefore,
regarding $Z$ as unknown root, we can certainly get the same solution
with Eq.\eqref{eq:ConstraintOneSecond}, i.e., Eq.\eqref{eq:RootZ}.
Substituting Eq.\eqref{eq:SolutionOfRadialSch} and Eq.\eqref{eq:SeriesN}
into Eq.\eqref{eq:TransformationRU}, we can also get the radial part
of the wave function,\begin{equation}
R_{n=2}=r^{|m|}e^{-\frac{1}{2}\omega_{L}r^{2}-\frac{k}{\omega_{L}}r}[\frac{(|m|+\frac{1}{2})\frac{k}{\omega_{L}}-Z}{|m|+\frac{1}{2}}r+1]a_{0}.\label{eq:RadialPartWaveFunctionNTwo}\end{equation}
 Substituting Eq.\eqref{eq:RootZ} into the above Eq.\eqref{eq:RadialPartWaveFunctionNTwo},
we get\[
R_{n=2}^{\pm}=r^{|m|}e^{-\frac{1}{2}\omega_{L}r^{2}-\frac{k}{\omega_{L}}r}[1-\frac{\frac{1}{2}\frac{k}{\omega_{L}}\pm\frac{1}{\omega_{L}}\sqrt{M}}{|m|+\frac{1}{2}}r]a_{0},\]
 which is coincide with Eq.\eqref{eq:RadialPartWaveFunctionJOneSecondPositive}
and Eq.\eqref{eq:RadialPartWaveFunctionJOneSecondNegative}.

4)$n\geq2$\begin{equation}
a_{n+1}=\frac{\omega_{L}(n+|m|+m)-\frac{1}{2}\frac{k^{2}}{\omega_{L}^{2}}-E}{(n+1)(|m|+\frac{1+n}{2})}a_{n-1}+\frac{(|m|+\frac{1}{2}+n)\frac{k}{\omega_{L}}-Z}{(n+1)(|m|+\frac{1+n}{2})}a_{n}.\label{eq:nArbitrary}\end{equation}
 Following the similar analysis above, also conferring to \cite{Taut1995Two},
it is a sufficient condition to guarantee normalizability of the eigenfunctions,
i.e.,\begin{equation}
E_{n}=\omega_{L}(n+|m|+m)-\frac{1}{2}\frac{k^{2}}{\omega_{L}^{2}},\label{eq:EnergyN}\end{equation}
 and\[
a_{n}=F(|m|,n,E,\omega_{L},k,Z)a_{0}=0.\]
 For $a_{0}\neq0$, we can deduce\begin{equation}
F(|m|,n,E,\omega_{L},k,Z)=0.\label{eq:F}\end{equation}
 Substituting Eq.\eqref{eq:EnergyN}into above Eq.\eqref{eq:F}, we
obtain\begin{equation}
F(|m|,n,\omega_{L}(n+|m|+m)-\frac{1}{2}\frac{k^{2}}{\omega_{L}^{2}},\omega_{L},k,Z)=0.\label{eq:RefinedF}\end{equation}
 From it, we obtain a constraint for $\omega_{L}$ and $k$, in another
word, $\omega_{L}$ and $k$ can't be chosen arbitrarily. Furthermore,
from Eq.\eqref{eq:nZero}, Eq.\eqref{eq:nOne} and Eq.\eqref{eq:nArbitrary},
we can get the polynomials of $\varphi$, i.e. Eq.\eqref{eq:SeriesN}.
By use of Eq.\eqref{eq:TransformationRU} and Eq.\eqref{eq:SolutionOfRadialSch},
we can get the corresponding $R_{n}.$

In conclusion, we have solved the Eq.\eqref{eq:radialsch} by an analytical
method. In the following section, we will build the connection between
the radial part equation and the sextic oscillator with a centrifugal
potential.

\section{KS Transformation}

\label{sec:ks}

Via KS transformation, i.e.,\begin{equation}
\begin{cases}
r=\rho^{2}\\
\theta=2\varphi\end{cases},\label{eq:KsTransformation}\end{equation}
 the Eq.\eqref{eq:TransformedSch} becomes\[
\frac{1}{2}[-\frac{1}{4\rho^{2}}(\frac{\partial^{2}}{\partial\rho^{2}}+\frac{1}{\rho}\frac{\partial}{\partial\rho}+\frac{1}{\rho^{2}}\frac{\partial^{2}}{\partial\varphi^{2}})\psi-i\omega_{L}\frac{\partial}{\partial\varphi}\psi+\omega_{L}^{2}\rho^{4}\psi]-\frac{Z}{\rho^{2}}\psi+k\rho^{2}\psi=E\psi.\]
 Substituting\begin{equation}
\psi(\rho,\varphi)=\frac{1}{\sqrt{2\pi}}e^{i\tilde{m}\varphi}\chi(\rho)\label{eq:KsWavefunction}\end{equation}
 into the above equation, where $\tilde{m}=0,\pm1,\pm2,...$, we get\[
-\frac{1}{2}(\frac{d^{2}}{d\rho^{2}}+\frac{1}{\rho}\frac{d}{d\rho}-\frac{\tilde{m}^{2}}{\rho^{2}})\chi+(2\tilde{m}\omega_{L}-4E)\rho^{2}\chi+4k\rho^{4}\psi+2\omega_{L}^{2}\rho^{6}\chi=4Z\chi.\]
 In order to cancel the item which contains $\frac{1}{\rho}$, substituting
\begin{equation}
\chi=\rho^{-\frac{1}{2}}\zeta\label{eq:KsRadialWaveFunctionTransformation}\end{equation}
 into the above equation, one obtains \begin{equation}
-\frac{1}{2}\frac{d^{2}\zeta}{d\rho^{2}}+\frac{4\tilde{m}^{2}-1}{8}\frac{1}{\rho^{2}}\zeta+(2\omega_{L}\tilde{m}-4E)\rho^{2}\zeta+4k\rho^{4}\psi+2\omega_{L}^{2}\rho^{6}\zeta=4Z\zeta,\label{eq:SexticSchrodinger}\end{equation}
 which is the so-called sextic oscillator with a centrifugal barrier
\cite{Levai2004sextic}. By comparison between Eq.\eqref{eq:TotalWaveFunction}
and Eq.\eqref{eq:KsWavefunction}, we can get\begin{equation}
m=\frac{\tilde{m}}{2},\label{eq:mKSm}\end{equation}
 and\begin{equation}
\chi(\rho)=R(r).\label{eq:KsRadialWaveFunctionRadialWF}\end{equation}
 Substituting Eq.\eqref{eq:KsRadialWaveFunctionTransformation} into
the above Eq.\eqref{eq:KsRadialWaveFunctionRadialWF}, we obtain\begin{equation}
\zeta=\sqrt{\rho}R(r),\label{eq:WaveFunctionOfSexticOscillator}\end{equation}
 which is precisely the wave function of the sextic oscillator with
a centrifugal barrier. Moreover, using Eq.\eqref{eq:KsTransformation}
and Eq.\eqref{eq:mKSm}, we can replace the parameters in Eq.\eqref{eq:WaveFunctionOfSexticOscillator}
by the counterparts used in Eq.\eqref{eq:SexticSchrodinger}. In section
two and three, we have calculate $Z$ by two different methods. Therefore,
it is very easy to get the eigenvalue $4Z$ of Eq.\eqref{eq:SexticSchrodinger}.

\section{Conclusion and acknowledgments}

\label{sec:conc}

In this paper, we have researched the two-dimensional hydrogen with
a linear potential in a homogeneous magnetic field by two different
methods, i.e., lie algebraic and analytic methods respectively. Furthermore
we have also build the connection between the radial part of the schr$\ddot{o}$dinger
equation and the sextic oscillator with centrifugal barrier.

This work was supported in part by NSF of China(Grants No.10605013
and No.10975075).

\end{document}